\documentclass[11pt]{article}
\usepackage[utf8]{inputenc}
\usepackage[T1]{fontenc}
\usepackage{hyperref}
\usepackage{url}
\usepackage{booktabs}
\usepackage{amsfonts}
\usepackage{amsmath}
\usepackage{nicefrac}
\usepackage{microtype}
\usepackage{graphicx}
\usepackage{natbib}
\usepackage{doi}
\usepackage[margin=1in]{geometry}
\usepackage{multirow}

\title{CytoDINO: Risk-Aware and Biologically-Informed Adaptation of \\
DINOv3 for Bone Marrow Cytomorphology}

\author{
  Aziz Muminov$^1$ \\
  \small muminova@dickinson.edu \\
  Anne Pham$^1$ \\
  \small phamn@dickinson.edu \\
  $^1$Department of Computer Science and Mathematics, Dickinson College, Carlisle, PA 17013
}

\date{}

\begin{document}
\maketitle

\begin{abstract}
Bone marrow cell cytomorphology analysis is critical for the diagnosis of hematological malignancies but remains a labor-intensive process subject to significant inter-observer variability. While recent foundation models have shown promise in computational pathology, they often require extensive computational resources and fail to account for the asymmetric risks associated with clinical misdiagnosis. We introduce CytoDINO, a framework that achieves state-of-the-art performance on the Munich Leukemia Laboratory (MLL) dataset by fine-tuning DINOv3 using Low-Rank Adaptation (LoRA). Our primary contribution is a novel \textbf{Hierarchical Focal Loss with Critical Penalties}, which encodes biological relationships between cell lineages and explicitly penalizes clinically dangerous misclassifications (e.g., classifying blasts as normal cells). CytoDINO achieves an \textbf{88.2\%} weighted F1 score and \textbf{76.5\%} macro F1 on a held-out test set of 21 cell classes. By utilizing parameter-efficient fine-tuning with only 8\% trainable parameters on a single NVIDIA RTX 5080, we demonstrate that consumer-grade hardware can match specialized infrastructure. Furthermore, confidence-based selective prediction yields \textbf{99.5\% accuracy on 67\% of samples}, suggesting a viable pathway for clinical deployment where high-uncertainty cases are flagged for expert review.
\end{abstract}


\section{Introduction}

Hematological diagnosis relies heavily on the microscopic examination of bone marrow smears to identify conditions such as acute myeloid leukemia (AML) and myelodysplastic syndromes~\citep{arber2022international, khoury2022}. The standard of care involves the manual cytomorphological analysis of 200–500 individual cells per patient. This process is time-consuming and prone to both inter- and intra-observer variability~\citep{font2015interobserver}. While automated image analysis offers the potential for improved consistency and throughput, progress has been hindered by limited annotated datasets, extreme class imbalance, and the strict requirement that clinical decision support systems minimize dangerous false negatives.

Recent developments in self-supervised learning (SSL) have produced robust vision transformers, notably DINO~\citep{caron2021emerging} and DINOv2~\citep{oquab2023dinov2}. The recently released DINOv3~\citep{simeoni2025dinov3}, trained on 1.7 billion images, represents the current state-of-the-art in general-purpose visual representation. In the specific domain of hematology, models like DinoBloom~\citep{koch2024dinobloom} have demonstrated that training on domain-specific data (380,000 hematology images) outperforms generalist models. However, fine-tuning these large models often requires computational resources unavailable to many clinical research centers.

A persistent challenge in hematopathology is the "long-tail" distribution of cell types. In the MLL bone marrow dataset~\citep{matek2021highly}, segmented neutrophils account for nearly 30,000 samples, whereas diagnostically critical faggot cells (indicative of acute promyelocytic leukemia) are represented by fewer than 50 examples. This imbalance reflects biological reality: normal cells vastly outnumber malignant ones. Furthermore, standard loss functions treat all errors equally. Clinically, however, misclassifying a malignant blast as a normal cell is a critical error that may delay life-saving treatment, whereas the inverse error triggers a review with less immediate harm.

\subsection{Contributions}

We present CytoDINO (Cytomorphology + DINO), a method for efficient, clinically-aware bone marrow cell classification. Our contributions are:

\begin{itemize}
    \item \textbf{Hierarchical Focal Loss:} We introduce a loss function that combines focal weighting with biologically-informed label smoothing. By allocating 95\% of the smoothing mass within-lineage and applying a 3$\times$ penalty for malignant-to-normal errors, we guide the model toward biologically reasonable and clinically safer predictions.
    \item \textbf{Performance:} CytoDINO achieves \textbf{88.2\%} weighted F1 and \textbf{76.5\%} macro F1 on unseen test data, surpassing the DinoBloom-G linear probing baseline (84.9\% wF1) on the same dataset.
    \item \textbf{Resource Efficiency:} Utilizing Low-Rank Adaptation (LoRA) with a rank of 64, we fine-tune only $\sim$8\% of the total parameters on a single NVIDIA RTX 5080 consumer GPU. This demonstrates that competitive medical AI does not strictly require massive-scale compute clusters.
    \item \textbf{Deployment Viability:} We demonstrate that confidence-based filtering enables \textbf{99.5\% accuracy on 67\% of samples}, validating a "human-in-the-loop" workflow where the model handles clear cases and experts review uncertain ones.
\end{itemize}

\section{Related Work}

\subsection{Foundation Models in Computer Vision}
Self-supervised learning has shifted the paradigm in computer vision from supervised training on limited labels to pre-training on vast unlabelled corpora. DINO~\citep{caron2021emerging} utilized knowledge distillation to learn distinct features without supervision. DINOv2~\citep{oquab2023dinov2} scaled this approach to 142 million images, enabling strong zero-shot classification. DINOv3~\citep{simeoni2025dinov3} further scales to 1.7 billion images and introduces Gram anchoring to prevent feature collapse in dense prediction tasks. We leverage DINOv3 as a robust feature extractor, hypothesizing that its scale provides morphological features transferable to microscopy.

\subsection{Hematology Image Analysis}
Deep learning for bone marrow analysis was significantly advanced by \citet{matek2019human} and \citet{matek2021highly}, who released large annotated datasets and established baselines using ResNeXt architectures. These CNN-based approaches, however, often struggle with rare classes. More recently, DinoBloom~\citep{koch2024dinobloom} applied the foundation model paradigm to hematology, aggregating 13 datasets to train a domain-specific ViT. While effective, DinoBloom primarily relies on linear probing for downstream tasks, which may limit the model's ability to adapt to complex decision boundaries compared to full or parameter-efficient fine-tuning.

\subsection{Addressing Class Imbalance and Efficiency}
Medical imaging datasets frequently exhibit long-tailed distributions~\citep{johnson2019mimic}. Common mitigation strategies include resampling, cost-sensitive learning, and Focal Loss~\citep{lin2017focal}, which focuses training on hard examples. Label smoothing~\citep{szegedy2016rethinking} is often used to improve calibration. Our work extends these concepts by incorporating domain-specific biological hierarchies into the smoothing and penalty terms. To address computational constraints, we employ Low-Rank Adaptation (LoRA)~\citep{hu2021lora}, which freezes pre-trained weights and injects trainable rank-decomposition matrices, significantly reducing memory overhead.

\section{Methods}

\subsection{Dataset}
We used the Bone Marrow Cytomorphology (BMC) dataset~\citep{matek2021highly}, consisting of 171,373 single-cell images annotated by experts from 945 patients. Images were captured at 40$\times$ magnification with May-Gr\"{u}nwald-Giemsa staining and cropped to 250$\times$250 pixels. The dataset contains 21 classes with extreme imbalance (Table~\ref{tab:class_distribution}). We perform a stratified 80/20 train-test split to ensure representation of rare classes in the evaluation set.

\begin{figure}[htbp!]
    \centering
    \includegraphics[width=0.55\linewidth]{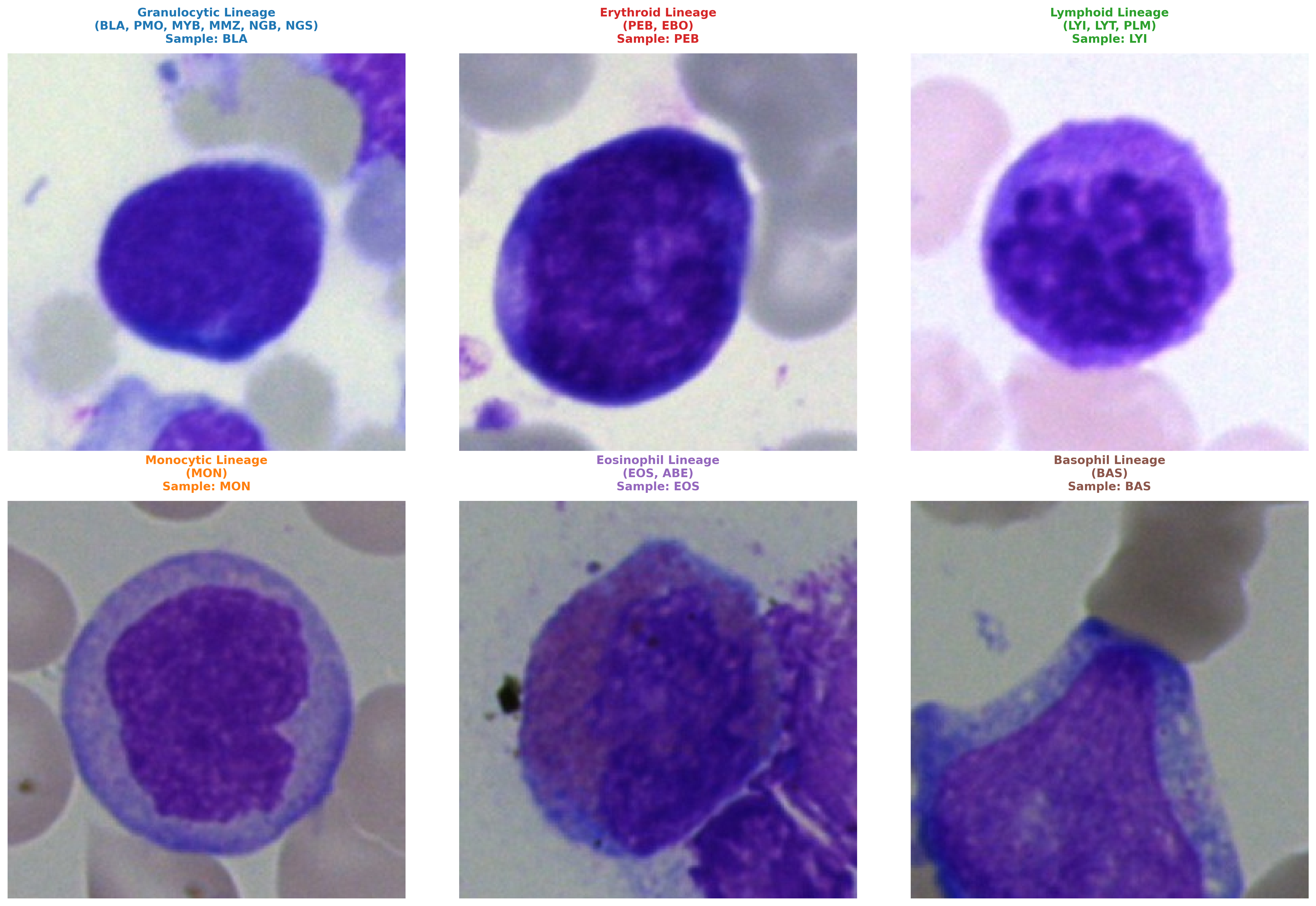}
    \caption{Representative bone marrow cells from six hematopoietic lineages showing distinct morphological characteristics. May-Grünwald-Giemsa staining, 40$\times$ magnification. Each lineage displays unique nuclear and cytoplasmic features that inform our hierarchical classification approach.}
    \label{fig:lineage_samples}
\end{figure}

Figure~\ref{fig:lineage_samples} illustrates representative cells from each of the six primary hematopoietic lineages in our dataset. The granulocytic lineage encompasses cells from early blasts through mature segmented neutrophils, representing the most diverse maturation sequence. The erythroid lineage contains proerythroblasts and erythroblasts at various maturation stages. Lymphoid cells include immature and mature lymphocytes as well as plasma cells. The monocytic lineage consists primarily of monocytes with characteristic irregular nuclei. Eosinophils and basophils, while rare in the dataset (Table~\ref{tab:class_distribution}), are identifiable by their distinctive cytoplasmic granulation patterns—coarse orange-red for eosinophils and deep purple-black for basophils. These morphological distinctions form the basis for our hierarchical loss formulation, which penalizes inter-lineage confusion more heavily than intra-lineage errors between adjacent maturation stages.

\begin{table}[h]
\centering
\caption{Class distribution demonstrating extreme imbalance in the BMC dataset.}
\label{tab:class_distribution}
\small
\begin{tabular}{@{}lr@{}}
\toprule
\textbf{Class} & \textbf{Samples} \\
\midrule
Segmented neutrophils (NGS) & 29,424 \\
Erythroblasts (EBO) & 27,395 \\
$\vdots$ & $\vdots$ \\
Basophils (BAS) & 441 \\
Faggot cells (FGC) & 47 \\
Abnormal eosinophils (ABE) & 8 \\
\bottomrule
\end{tabular}
\end{table}
\subsection{Data Balancing and Robustness Strategy}

Given the heavy imbalance in the BMC dataset (ranging from $\sim$30,000 samples for neutrophils to $<10$ for pathological variants), standard uniform sampling yields batches dominated by majority classes, causing gradient starvation for rare classes. We address this via a three-pronged approach:

\subsubsection{Stochastic Damped-Inverse Sampling}
We implemented a custom \texttt{BalancedOversampler} that constructs batches using multinomial sampling with replacement. To prevent overfitting on rare classes (a common risk with strict inverse-frequency sampling), we employed a damped weighting heuristic. The sampling probability $P(x)$ for an image belonging to class $c$ is defined as:

\begin{equation}
P(x \in c) \propto \frac{1}{(N_c + 1)^{\lambda}}
\end{equation}

where $N_c$ is the cardinality of class $c$ and $\lambda=0.7$ is a smoothing factor. This hyperparameter $\lambda$ strikes a balance: it up-samples rare pathologies (e.g., Faggot cells) sufficiently to stabilize training dynamics, while retaining enough density of the majority classes (e.g., Segmented Neutrophils) to learn their diverse morphological variations.

\subsubsection{Epoch Expansion}
To accommodate the oversampling of rare classes without discarding unique majority samples, we defined a training epoch as $2\times$ the cardinality of the original training set. This expanded epoch length ensures that while rare samples are repeated multiple times per epoch to reinforce feature learning, the model is still exposed to the full breadth of variance present in the majority classes.
\subsubsection{Augmentation-Driven Regularization}
The oversampling strategy described above introduces a risk of overfitting, whereby the model may memorize texture artifacts or staining idiosyncrasies rather than learning generalizable morphological features. To mitigate this, we implemented an extensive augmentation pipeline using Albumentations~\citep{info2020albumentations}, designed to increase effective sample diversity while preserving clinically relevant cell characteristics.

\begin{figure}[htbp!]
    \centering
    \includegraphics[width=0.55\linewidth]{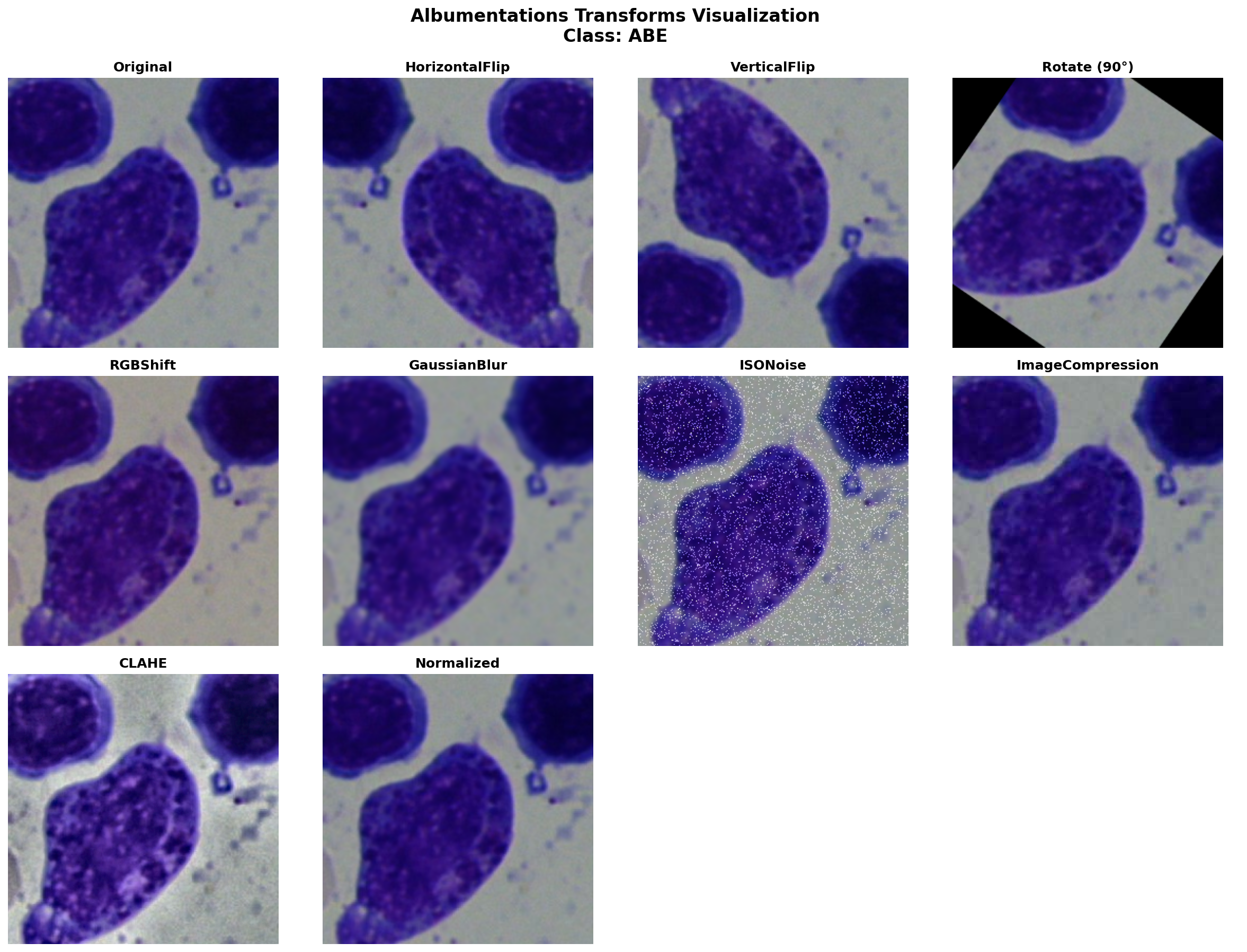}
    \caption{Augmentation pipeline applied during training. Representative transformations include geometric operations (flips, rotations), photometric perturbations (color jitter, brightness/contrast adjustments), and noise injection (Gaussian blur, ISO noise). The rightmost panel shows the final normalized image using dataset-specific statistics ($\mu = [0.5631, 0.4959, 0.7355]$, $\sigma = [0.2419, 0.2835, 0.1761]$) computed from the entire MLL dataset. These transformations simulate inter-laboratory variability in staining protocols and imaging equipment while maintaining biological plausibility.}
    \label{fig:augmentation_pipeline}
\end{figure}

Our augmentation strategy (Figure~\ref{fig:augmentation_pipeline}) comprises two complementary components:

\begin{itemize}
    \item \textbf{Geometric Transformations:} We applied random horizontal and vertical flips along with rotations uniformly sampled from $[0^\circ, 360^\circ]$. This exploits the inherent rotational and reflectional symmetry of suspended cells in bone marrow smears, where orientation carries no diagnostic significance.
    
    \item \textbf{Photometric Perturbations:} To enhance robustness to inter-laboratory variations in staining protocols and digitization hardware, we applied stochastic color augmentations including brightness ($\pm10\%$), contrast ($\pm10\%$), saturation ($\pm15\%$), and hue shifts ($\pm3\%$). Additionally, we introduced sensor-level noise through Gaussian blur (kernel size 3--5) and ISO noise (intensity 0.1--0.5) with probability 0.3, simulating variations in microscope optics and camera sensor characteristics across different institutions.
\end{itemize}

All augmented images were subsequently normalized using channel-wise mean and standard deviation statistics computed from the complete MLL training set ($\mu_{RGB} = [0.5631, 0.4959, 0.7355]$, $\sigma_{RGB} = [0.2419, 0.2835, 0.1761]$). This normalization was applied consistently to both training and validation sets to ensure distribution alignment. By combining aggressive augmentation with careful normalization, we enable the model to focus on morphological features invariant to technical variation while maintaining sensitivity to diagnostically relevant cellular characteristics.
\subsection{Model Architecture}

\subsubsection{DINOv3 Backbone with LoRA}
We employ the DINOv3-ViT-L backbone ($\sim$304M parameters). To preserve the generalized feature space learned during pre-training, we freeze the backbone weights and inject LoRA adapters into the query, key, value, and output projection layers of the attention blocks. The forward pass for a linear layer with weight $W_0$ is modified as:
\begin{equation}
W'x = W_0 x + \alpha \cdot B A x
\end{equation}
where $W_0$ is frozen, $B \in \mathbb{R}^{d \times r}$ and $A \in \mathbb{R}^{r \times k}$ are trainable matrices with rank $r=64$, and $\alpha=128$ is a scaling factor.

\subsubsection{Transformer Decoder Head}
Standard linear probing utilizes only the class token (CLS). To better capture morphological details distributed across the cell image, we implement a lightweight transformer decoder. This decoder aggregates information from all patch embeddings $\mathbf{z}$:
\begin{align}
\mathbf{q} &\in \mathbb{R}^{1 \times d} \quad \text{(learnable query token)} \\
\mathbf{h} &= \text{TransformerDecoder}(\text{query}=\mathbf{q}, \text{memory}=\mathbf{z}) \\
\hat{y} &= \text{Linear}(\text{LayerNorm}(\mathbf{h}))
\end{align}
The resulting architecture contains $\sim$25M trainable parameters, representing roughly 8\% of the total model size.

\subsection{Hierarchical Focal Loss with Critical Penalties}
We propose a loss function $L$ designed to embed hematological domain knowledge directly into the optimization process.

\subsubsection{Biologically-Informed Label Smoothing}
Standard label smoothing distributes probability mass uniformly across non-target classes. We instead construct a smoothing matrix $S$ that respects hematopoietic lineages (e.g., Granulocytic, Erythroid, Monocytic). We define the target distribution as:
\begin{equation}
S[i,j] = \begin{cases}
1 - \epsilon & \text{if } j = i \\
\epsilon \cdot w_{\text{within}} \cdot \mathbf{1}[j \in L_i] / |L_i| & \text{if } j \in L_i \text{ (same lineage)} \\
\epsilon \cdot w_{\text{cross}} / |C| & \text{otherwise}
\end{cases}
\end{equation}
With $\epsilon=0.1$ and $w_{\text{within}}=0.95$, the model is penalized less for confusing a Promyelocyte with a Myelocyte (same lineage) than for confusing it with a Lymphocyte.

\subsubsection{Critical Error Penalty}
We define a penalty matrix $M$ to distinguish between safe and dangerous errors. Let $\mathcal{C}$ represent critical classes (malignant/immature, e.g., Blasts, Faggot cells) and $\mathcal{S}$ represent safe classes (mature/normal).
\begin{equation}
M[i,j] = \begin{cases}
3.0 & \text{if } i \in \mathcal{C}, j \in \mathcal{S} \text{ (False Negative / Dangerous)} \\
1.5 & \text{if } i \in \mathcal{S}, j \in \mathcal{C} \text{ (False Positive)} \\
1.0 & \text{otherwise}
\end{cases}
\end{equation}

\subsubsection{Combined Objective}
The final loss combines the penalty matrix, class-balanced weights $\alpha$, Focal Loss to address easy/hard samples, and the smoothed targets:
\begin{equation}
\mathcal{L} = \frac{1}{N} \sum_{i=1}^{N} M[y_i, \hat{y}_i] \cdot \alpha_{y_i} \cdot (1-p_{t,i})^\gamma \cdot \text{CE}(S[y_i], \hat{p}_i)
\end{equation}
where $\gamma=2.0$.

\subsection{Training Configuration}
Training was conducted on a single NVIDIA GeForce RTX 5080 (16GB GDDR7). We optimized using AdamW with a learning rate of $3\times10^{-4}$ and cosine decay. The batch size was set to 16 with gradient accumulation steps of 4, enabling an effective batch size of 64. Training concluded after 8 epochs.

\section{Results}

\subsection{Comparative Performance}
Table~\ref{tab:main_results} presents the performance of CytoDINO against current state-of-the-art methods on the BMC dataset. CytoDINO outperforms the DinoBloom-G foundation model (linear probe baseline) across all aggregate metrics. Notably, we achieve a Weighted F1 score of 88.2\%, improving upon the previous best of 85.7\% (DinoBloom-S).

\begin{table}[h]
\centering
\caption{Performance comparison on BMC dataset (21 classes). CytoDINO utilizes LoRA fine-tuning, while baselines use linear probing.}
\label{tab:main_results}
\begin{tabular}{@{}lccc@{}}
\toprule
\textbf{Model} & \textbf{Weighted F1} & \textbf{Macro F1} & \textbf{Bal. Acc.} \\
\midrule
DINOv2 ViT-G & 73.5\% & -- & 52.1\% \\
CTransPath & 74.1\% & -- & 52.2\% \\
Phikon ViT-B & 73.2\% & -- & 54.4\% \\
\midrule
DinoBloom-S & 85.7\% & -- & 71.4\% \\
DinoBloom-L & 84.9\% & -- & 64.4\% \\
DinoBloom-G & 84.9\% & -- & 69.3\% \\
\midrule
\textbf{CytoDINO (Ours)} & \textbf{88.2\%} & \textbf{76.5\%} & \textbf{75.7\%} \\
\bottomrule
\end{tabular}
\end{table}

\subsection{Class-Specific Analysis}
Table~\ref{tab:per_class} details performance on representative classes. The model shows robust performance on high-frequency classes (Erythroblasts: 95.1\% F1). More importantly, it achieves high sensitivity on rare, pathological classes. Faggot cells, critical for the diagnosis of acute promyelocytic leukemia, are identified with an F1 score of 82.4\% despite there being only 9 examples in the test set. Similarly, Smudge cells (CLL marker) are detected with 87.5\% F1.

\begin{table}[h]
\centering
\caption{Per-class performance on test set (selected classes).}
\label{tab:per_class}
\small
\begin{tabular}{@{}lrrrr@{}}
\toprule
\textbf{Class} & \textbf{Support} & \textbf{Prec.} & \textbf{Rec.} & \textbf{F1} \\
\midrule
Seg. neutrophils (NGS) & 5,884 & 93.3 & 89.9 & 91.5 \\
Erythroblasts (EBO) & 5,479 & 95.4 & 94.8 & 95.1 \\
Blasts (BLA) & 2,394 & 88.2 & 85.8 & 87.0 \\
\midrule
Hairy cells (HAC) & 81 & 82.5 & 81.5 & 82.0 \\
Faggot cells (FGC) & 9 & 87.5 & 77.8 & 82.4 \\
Smudge cells (KSC) & 8 & 87.5 & 87.5 & 87.5 \\
\bottomrule
\end{tabular}
\end{table}

\subsection{Error Topology}
An analysis of confusion matrices reveals that the top five most frequent misclassifications (Table~\ref{tab:confusions}) occur exclusively between adjacent maturation stages within the same lineage (e.g., Segmented Neutrophils vs. Band Neutrophils; Myelocytes vs. Promyelocytes). This confirms that the hierarchical smoothing successfully constraints the model to make "biologically plausible" errors rather than random across-lineage mistakes.

\begin{table}[h]
\centering
\caption{Top confusion pairs on the test set.}
\label{tab:confusions}
\begin{tabular}{@{}llrl@{}}
\toprule
\textbf{True} & \textbf{Pred.} & \textbf{N} & \textbf{Relationship} \\
\midrule
NGS & NGB & 442 & Same lineage, adjacent \\
NGB & NGS & 241 & Same lineage, adjacent \\
MYB & PMO & 214 & Same lineage, adjacent \\
PMO & MYB & 192 & Same lineage, adjacent \\
\bottomrule
\end{tabular}
\end{table}

\subsection{Selective Prediction for Clinical Workflows}
To assess clinical utility, we evaluated a selective prediction strategy where predictions below a softmax confidence threshold are flagged for human review. As shown in Table~\ref{tab:coverage}, setting a threshold of 95\% yields a subset of predictions (66.7\% of the data) with \textbf{99.5\% accuracy}. This implies that in a clinical setting, two-thirds of cells could be auto-classified with near-perfect reliability, significantly reducing the manual workload.

\begin{table}[h]
\centering
\caption{Accuracy vs. Coverage at varying confidence thresholds.}
\label{tab:coverage}
\begin{tabular}{@{}cccc@{}}
\toprule
\textbf{Threshold} & \textbf{Coverage} & \textbf{Accuracy} & \textbf{Flagged} \\
\midrule
70\% & 93.3\% & 98.2\% & 6.7\% \\
90\% & 78.5\% & 99.2\% & 21.5\% \\
\textbf{95\%} & \textbf{66.7\%} & \textbf{99.5\%} & \textbf{33.3\%} \\
\bottomrule
\end{tabular}
\end{table}

\subsection{Learned Representation Analysis}
\begin{figure}[htbp!]
    \centering
    \includegraphics[width=1.0\linewidth]{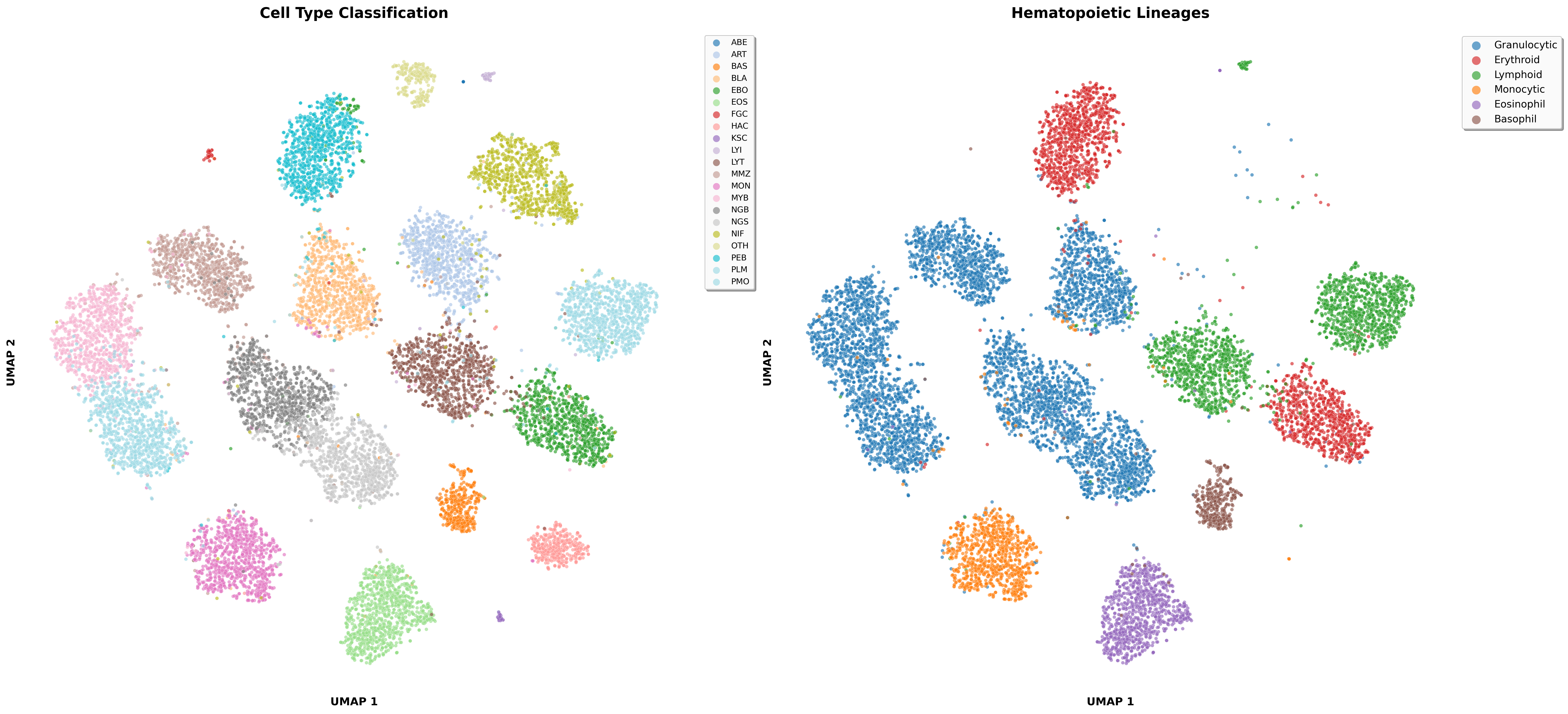}
    \caption{TSNE visualization of learned embeddings from the Transformer Decoder Head. \textbf{Left:} Embeddings colored by cell class (21 classes) show clear separation with minimal overlap. \textbf{Right:} Embeddings colored by lineage reveal hierarchical organization, with the granulocytic lineage (blue) forming a continuous trajectory from immature blasts to mature neutrophils. The model learns biologically meaningful representations that cluster related cell types while maintaining discriminative power for fine-grained classification.}
    \label{fig:embedding_overview}
\end{figure}

To understand the quality of learned features, we visualized the 1024-dimensional embeddings produced by our Transformer Decoder Head using TSNE dimensionality reduction~\citep{van2008visualizing}. Figure~\ref{fig:embedding_overview} presents two complementary views of the embedding space computed from the validation set. The left panel shows embeddings colored by individual cell class, revealing clear clustering of morphologically similar cell types. The right panel groups cells by hematopoietic lineage, demonstrating that CytoDINO learns a hierarchically organized representation space that respects biological taxonomy.

Notably, the granulocytic lineage forms a continuous manifold structure reflecting the maturation sequence (BLA → PMO → MYB → MMZ → NGB → NGS), suggesting the model captures developmental trajectories rather than treating each class independently. Lineages are well-separated in embedding space, with minimal cross-lineage overlap, confirming that our Hierarchical Focal Loss successfully encouraged biologically coherent representations. This structured embedding space explains why most classification errors occur between adjacent maturation stages within the same lineage (Table~\ref{tab:confusions}) rather than across unrelated cell types.

\subsection{Zero-Shot Generalization}
We further evaluated robustness by applying the model trained on MLL bone marrow data directly to the RAABIN-WBC peripheral blood dataset~\citep{kouzehkanan2022large} without fine-tuning. Despite substantial domain shifts—including different staining protocols (Wright-Giemsa vs. May-Grünwald-Giemsa), cell types (peripheral blood vs. bone marrow), and digitization hardware—CytoDINO achieved \textbf{86.3\% accuracy} on RAABIN-WBC test set A. This was accomplished by mapping our 21 fine-grained bone marrow classes to RAABIN's 5 coarse white blood cell categories (neutrophils, lymphocytes, monocytes, eosinophils, basophils) via probabilistic aggregation of class posteriors.

The strong zero-shot performance indicates that CytoDINO learns generalizable morphological features rather than memorizing dataset-specific staining artifacts or scanner characteristics. This cross-dataset transferability is particularly encouraging for clinical deployment, where models must handle variations in sample preparation and imaging equipment across different laboratories.

\section{Discussion}

The performance of CytoDINO highlights the efficacy of combining general-purpose foundation models (DINOv3) with domain-specific constraints. The Hierarchical Focal Loss function proved critical in two regards: first, by clustering errors within biological lineages, it ensures that misclassifications retain diagnostic utility (e.g., identifying a cell as granulocytic, even if the maturation stage is inexact). Second, the penalty matrix successfully mitigated critical errors, maintaining high sensitivity for rare malignancies like Faggot cells despite extreme class imbalance.

Limitations of this study include the extremely low sample size for certain rare classes (e.g., Abnormal Eosinophils), which prevents definitive statistical conclusions for those specific subtypes. Additionally, while the zero-shot evaluation on RAABIN is promising, multi-center validation on bone marrow data is required to fully confirm robustness against stain variations.

\section{Conclusion}

We introduced CytoDINO, a parameter-efficient framework for bone marrow cell classification. By fine-tuning DINOv3 with a Hierarchical Focal Loss, we achieved state-of-the-art results (88.2\% Weighted F1) while utilizing accessible consumer hardware. The model demonstrates biologically reasonable error patterns and supports a high-confidence selective prediction workflow suitable for clinical screening. Future work will focus on integrating these cell-level predictions into patient-level diagnostic models for leukemia sub-typing.

\bibliographystyle{unsrtnat}

\end{document}